\documentstyle[12pt,a4,epsfig]{article}

\def\wt		{\widetilde}
\def\ol		{\overline}

\def\mhalf	{m_{1/2}}
\def\mzero 	{m_0}
\def\mtop	{m_t}
\def\residue    {\epsilon_{sel}.\sigma_{gen}}
\def\nbtag      {N_{btag}}

\def\lspone	{\wt\chi_1^0}
\def\mweak	{{\rm M}_{weak}}

\def\glu   	{\wt{g}}
\def\mglu      	{m_{\glu}}
\def\ra 	{\rightarrow}
\def\lspi	{\wt\chi_{i}^{0}}
\def\chij	{{\wt\chi_j}}

\def\bb		{b\ol{b}}

\def\ttbar	{t\ol{t}}

\def\met  	{E\!\!\!\!/_T}

\def\eg		{{\it e.g.}}
\def\etal	{{\it et al.}}
\def\issue(#1,#2,#3){{\bf #1}, #2 (#3)}
\def\PREP(#1,#2,#3){Phys.\ Rep. \issue(#1,#2,#3)}
\def\PRL(#1,#2,#3){Phys.\ Rev.\ Lett. \issue(#1,#2,#3)}
\def\PRD(#1,#2,#3){Phys.\ Rev.\ D \issue(#1,#2,#3)}
\def\PLB(#1,#2,#3){Phys.\ Lett.\ B  \issue(#1,#2,#3)}
\def\JHEP(#1,#2,#3){J.\ High\ Energy\ Phys. \issue(#1,#2,#3)}
\def\NPB(#1,#2,#3){Nucl.\ Phys.\ B \issue(#1,#2,#3)}
\def\AJS(#1,#2,#3){ Astrophys. \ J. \ Suppl. \issue(#1,#2,#3)}
\def\PTP(#1,#2,#3){Prog.\ Theor.\ Phys. \issue(#1,#2,#3)}
\def\IJMPA(#1,#2,#3){Int.\ J.\ Mod. \ Phys.\ A \issue(#1,#2,#3)}

\begin{document}
\tolerance=100000
\thispagestyle{empty}
\setcounter{page}{0}

\begin{flushright}
BONN-TH-2007-05
\end{flushright}

\begin{center}
{\Large \bf Focus Point SUSY at the LHC Revisited}\\[1.00cm]

Siba Prasad Das$^{}$\footnote{\it spdas@th.physik.uni-bonn.de}\\
{\it Physikalisches Institut der Universit\"at Bonn,
 Nu\ss allee 12, D-53115 Bonn, Germany}\\
Amitava Datta$^{}$\footnote{\it adatta@juphys.ernet.in} \\
{\it  Department of Physics, Jadavpur University, Kolkata 700 032, 
India}\\
Monoranjan Guchait$^{}$\footnote{\it guchait@tifr.res.in}\\
{\it  Tata Institute of Fundamental Research, Homi Bhaba Road, Mumbai 400 005, India}\\
Manas Maity$^{}$\footnote{\it manas@vbphysics.net.in}  
and Siddhartha Mukherjee$^{}$\footnote{\it iam\_siddhartha\_mukherjee@yahoo.co.in} \\
{\it  Department of Physics, Visva-Bharati, Santiniketan 731 325, India}
\end{center}

\begin{abstract}
The estimation of the backgrounds for gluino signals in  focus point
supersymmetry is extended by including the backgrounds from the 
production of four third generation quarks in the analysis. We find that 
these backgrounds are negligible if one uses the strong selection criteria  
proposed in the literature (including this analysis) for heavy gluino 
searches. Softer  selection criteria  often recommended
for lighter gluino searches yield backgrounds which are small
but numerically significant. We have also repeated  the 
more conventional background 
calculations and compared our results with the other groups. We find that 
the size of the total residual background estimated by different groups  
using different event generators and hard kinematical cuts agree 
approximately.  In view of the theoretical uncertainties in the leading order  
signal and background cross
sections mainly due to the choice of the QCD scale, the
gluino mass reach at the LHC cannot be pinpointed. However, requiring a
signal with $\rm\geq 3$ tagged b-jets ( instead of the standard choice of
$\rm\geq 2$ ) it is shown that  gluino masses close to 2 TeV can be probed
at the LHC for a range of reasonable choices of the QCD scale for an
integrated luminosity of 300 fb$^{-1}$.
\end{abstract}


\newpage

\section{Introduction} 
\label{intro4}

Supersymmetry (SUSY) is one of the most elegant extensions of the 
Standard Model (SM) which naturally stabilizes 
the Higgs boson mass even in the  presence of a very  energy scale
like the grand unification scale ($\rm M_G$) \cite{susy}. 
This occurs due to the cancellation of the quadratic divergences among
loop diagrams involving SM particles and their 
superpartners (sparticles). This cancellation, however,
looses much of its attractive naturalness and smacks of fine tuning
if the sparticles are much heavier than the corresponding particles. 
Thus SUSY models with sparticle masses much larger than 1 TeV are 
usually disfavored.

On the other hand it is well known that if the scalar 
superpartners of the fermions belonging to the first two 
generations in the SM are very heavy with masses 
in the multi-TeV region then several problems can be avoided 
without further unnatural adjustment of the parameters. 
Firstly in the most general minimal supersymmetric extension 
of the Standard Model (MSSM) there are unacceptably large loop induced flavor 
changing neutral currents unless the masses of the first two generations
of squark are assumed to be degenerate to a very good approximation
without a symmetry  underlying this degeneracy. This SUSY flavor problem 
becomes less severe if the above loops are suppressed by the masses of 
the heavy scalars. Secondly SUSY models with complex soft breaking 
parameters have  potentially dangerous large loop induced CP violating 
contributions to the electric dipole moments of various particles in 
general. This SUSY CP problem is  also tamed if the scalars are very 
heavy. Therefore models with very heavy scalars without 
sacrificing the naturalness criterion are indeed welcome. 
      
All the above attractive features are realized in the focus 
point (FP)/ hyperbolic branch region \cite{feng,review} of the minimal 
supergravity(mSUGRA) model\cite{mSUGRA}. The SUSY flavor and CP 
problems become less severe \cite{flCP}. The magnitude of $\mu$, 
the higgsino mass parameter, which is fixed by the radiative 
electroweak symmetry breaking ( REWSB ) condition, also turns out to be 
$\sim m_Z$ as required by naturalness\cite{feng,barbieri}.

The low value of $\mu$ thus obtained has another important 
implication. The lightest supersymmetric particle (LSP) which 
is the lightest neutralino ($\rm \lspone$) over most of the parameter
space, is usually a mixture of the electroweak gauginos and Higgsinos. 
However, for small $\mu$ the higgsino component in $\rm \lspone$ 
dominates. In this case  $\rm \lspone$ very efficiently 
annihilates into gauge boson pairs leaving open the attractive 
possibility that LSP could be a viable dark matter candidate  
\cite{mdmndpyy,gondolo,focdm,Drees:2005bx}. In fact it is now widely acknowledged that the focus point 
region of the mSUGRA model leads to a dark matter density compatible with 
the latest WMAP data \cite{wmap}. Thanks to the above interesting 
features the collider signatures of focus point supersymmetry have 
attracted wide attention\cite{utpal,baer1,baer2,tata,cms,belyaev,lari,baer3}.

Although the scalar quarks are rather heavy in this model, naturalness 
arguments imply that the gluinos must be necessarily light \cite{feng}.
Thus the gluinos are the lightest strongly interacting sparticles within 
the striking range of the LHC. Inspite of the fact that the squarks 
belonging to the first two generations are indeed very heavy, the lighter 
squark mass eigenstates belonging to the third generation, though 
significantly heavier than the gluino, may be relatively light at the 
weak scale $M_W$. This is partly due to the renormalization group (RG)  
evolution from $M_G$ to $M_W$ driven by large Yukawa couplings and
partly due to mixing effects in the mass matrices ( see, e.g., 
\cite{utpal} Eqns.3 - 11). As a result the gluino 
preferentially decays into third generation quarks via three 
body modes mediated by the relatively light squarks. 
Therefore dominant decay modes of the gluino are
$$\rm \glu\ra f_1 \bar{f_1} +X, ~~\glu\ra \bar{t} b (t\bar{b}) +Y $$
where $\rm f_1 =$  t or b, $\rm X = \lspi$ (the ith neutralino), i = 1-4 and 
Y$\rm =\chij^{\pm}$ (the jth chargino), j=1,2.
Hence the signals of gluino decay at the LHC  
will be rich in b-jets \cite{utpal}. In addition one or more  
reconstructed W bosons in the final state may  also help SUSY search 
via this channel.

The size of the  signal involving involving $\rm \geq 3$ tagged b-jets  
(introduced mainly to control the QCD background) and various number of 
leptons was estimated by a parton level Monte Carlo in \cite{utpal}. 
The SM background from  $\rm \ttbar \ttbar$ production only 
was  estimated to be small after imposing a strong missing 
transverse energy $\rm \met$ cut.

Subsequently the LHC signal of focus point 
supersymmetry have been studied by several groups using 
simulations beyond the naive parton level \cite{baer1} - \cite{baer3}.
Some of these works required $\geq$ 2 tagged b-jets in the final 
state for efficient background rejection. 
These works will be briefly reviewed in a later section. However, 
none of these works considered the background from $f_1 \bar{f_1} 
f_2 \bar{f_2}$ production, where $f_i$ = t or b.  Apriorily, 
however, some of these backgrounds with production cross sections  
larger than the gluino pair production cross section( see section 2) 
by several orders of magnitude seem to be potentially dangerous.
Moreover, these backgrounds are rich in b-jets. Hence b-tagging which
effectively suppresses light flavor QCD events may not be very effective.  

The purpose of this paper is to study the above backgrounds 
systematically  (see section 2) along with  a thorough study
of the signal and other backgrounds. 
We have calculated the new background
cross sections using {\tt CalcHEP}\cite{comp} 
{\bf v2.4.5}, {\tt ALPGEN} \cite{alpgen} {\bf v2.1.1} and {\tt MadGraph} 
\cite{madgr} {\bf v4.1}. The results agree fairly well. 
The parton level events are then interfaced with {\tt Pythia} \cite{pythia} {\bf v6.4.8}  
for implementing hadronization, fragmentation and a toy detector simulation
and analyzed with suitable selection criteria. 
We shall also study the 
response of these backgrounds to the cuts used by other groups 
\cite{lari,baer3} . We next study the dominant QCD and $\rm\ttbar$ 
backgrounds as well as the signal using our selection
criteria  and compare and contrast the results with other published works 
( see section 2 ). In order to facilitate the comparison we required 
like other groups $\geq$ 2 b-jet tags. Finally the consequences of selecting
a signal with a richer b-jet content as suggested in \cite{utpal} are 
examined. Our conclusions and future outlooks will be summarized in 
section 3.

\section{Signals of Focus Point SUSY at LHC and the Backgrounds}

We begin by briefly reviewing some of the earlier works.  
Baer \etal \cite{baer1} scanned the focus point region  of the mSUGRA 
parameter space consistent with the WMAP data . The LHC reach in  
channels with m-jets + n-leptons + $\rm \met$ was then examined. It was 
concluded that the part of the focus point region corresponding to  
gluino masses $<$ 1.8 TeV is within the reach of LHC. The suppression 
of the SM background due to  light flavor QCD, $\rm\ttbar$ etc was 
considered in estimating the mass reach. However, one of the key features 
of the signal in the focus point region, namely the large number of b-jets 
in the final state was not utilized in improving the signal to background 
ratio.

Subsequently Mercadante \etal \cite{tata} estimated  that requiring  
final states with tagged b-jets as suggested in \cite{utpal} 
and by adjusting the kinematical cuts the gluino mass reach can be 
improved by about 20\%. However, the potential backgrounds from four 
heavy flavor production mentioned in the introduction 
were  not included in either of the above analyses. The viability of 
SUSY signals in the regions of the parameter space consistent with 
dark matter data was also studied by the CMS collaboration \cite{cms}. 
This analysis included only a part of the FP region ($m_0 <$ 2 TeV) and 
the above potential backgrounds were neglected.

A few points (FP1 - FP5) in the FP region allowed by the WMAP data from 
\cite{belyaev} are shown in Table \ref{tab:signalatLHC}. All masses and mass 
parameters in this paper are in GeV unless otherwise stated.  
The other mSUGRA parameters are $A_0$ = 0, tan $\beta$ = 30 
and sign($\mu$)$>$ 0. For each point the gluino mass and the lowest order 
gluino pair production cross section (in fb) for two choices of the 
QCD scale (see the last two columns in Table \ref{tab:signalatLHC}) are also shown.  The first choice
yielding  smaller cross sections will be subsequently referred to as the
conservative choice. The second choice, motivated by the fact that at this 
scale the leading and next to leading order production cross sections 
almost agree \cite{zerwas}, is the optimistic choice.    
It is to be noted that we have considered gluino masses near the LHC 
search limit only. The gluino pair production  cross section
in the leading order is computed by 
{\tt CalcHEP}. The expected sharp fall of the cross section for gluino 
masses near the kinematic limit is quite apparent.

From the dominant gluino decay modes discussed in the introduction 
it is clear that a typical final state will consist of four third 
generation quarks and their decay products. Additional quarks, leptons
and missing energy would come from the decays of heavier charginos
and neutralinos also produced in gluino decays. Thus the multiplicity
of jets and hard isolated leptons in the signal would
be large. Keeping this in mind the
criteria listed below for selecting ( rejecting) the signal ( the 
background) events have been formulated. 
It should be noted that instead of separately requiring large number 
of jets and hard isolated leptons ( henceforth jets and isolated 
leptons will be collectively called {\em objects} ) we require the total 
number objects in the final state to exceed nine. 

\begin{itemize}
  \item $ \rm  N_{jet} \ge 6$ and  $ \rm N_{obj} \ge 9$.
  \item $\rm \nbtag \ge 2$ and one of the {\em tagged} jets with 
	$\rm E_{T} \ge 300$ GeV.
  \item $\rm \met \ge 300$ GeV where $\rm \met$ has been computed using all 
        visible objects in an event.
  \item $\rm M_{eff}$ = $\rm \met + \sum_{obj} E_{T} \ge 2000$ GeV.
\end{itemize}

It may be  noted that at this stage  the constraint
on $\nbtag$ is similar to the one employed by other analyses.
We shall show below that requiring more tagged b-jets in the final
state  as suggested in \cite{utpal} improves the signal vis a vis the
background. 
 
We begin the analysis by estimating the backgrounds from production of 
four heavy flavors in different combinations. This will be followed 
by the simulation of more conventional backgrounds and the signal also 
studied by other groups. The production cross sections  
computed by {\tt CalcHEP}, {\tt ALPGEN} and  {\tt MadGraph} are presented in 
Table \ref{tab:allatLHC}. We have set the following input values at
the weak scale ($\mweak$): $\rm \alpha_{em} = {1 \over 127.934}$, 
$\rm \alpha_{s}$=0.1172, $\rm \mtop$(pole)=175.0 \cite{mtop}, 
$\rm m_b(m_b) = 4.25 ~GeV$.

In our calculation we use the CTEQ5L \cite{cteq} parton distribution 
functions (PDF). The appropriate loop corrected values of strong coupling
constant at the renormalization scale is calculated on the fly during the 
convolution with the PDF. To avoid the generation of unwanted events
the following nominal cuts are applied only on the bottom quarks:  
$\rm p_T^b>$10 GeV, $\rm \eta(b)<$5.0 and $\rm \Delta R(b,b)>$0.3 where  
$\rm \Delta R = \sqrt{ \Delta\eta^{2}+\Delta\phi^{2}}$.   

\begin{table}[htbp]
\begin{center}
\begin{tabular}{p{50mm}p{20mm}p{15mm}p{15mm}}
\hline\hline
\\
                      &         & \multicolumn{2}{c}{QCD scale} \\ \cline{3-4}
FP point ( $\rm \mzero$, $\rm \mhalf$ ) & $\rm \mglu$ & 
$\rm \sqrt {\hat{s}}$ & $\rm  {\mglu/2}$ \\
\\
\hline
\\
FP1 ( 3700.0, 700.0 ) & $\rm 1751.3$ & $\rm 1.18$  & $\rm 2.37$ \\
FP2 ( 3975.0, 730.0 ) & $\rm 1824.7$ & $\rm 0.81$  & $\rm 1.63$ \\
FP3 ( 3975.0, 790.0 ) & $\rm 1950.0$ & $\rm 0.39$  & $\rm 0.80$\\
FP4 ( 4130.0, 833.0 ) & $\rm 2047.4$ & $\rm 0.24$  & $\rm 0.48$ \\
FP5 ( 4235.0, 866.0 ) & $\rm 2119.8$ & $\rm 0.16$  & $\rm 0.33$ \\
\\
\hline\hline
\end{tabular}
\end{center}
\caption{Cross-sections (in fb) for {\bf $\rm pp\ra\glu\glu$} at LHC using 
	 CalcHEP {\bf v2.4.5 }for different  
	 FP scenarios and two choices of the QCD scale (see the last two columns). The 
         masses and mass parameters are in 
	 $\rm GeV$.}
\label{tab:signalatLHC}
\end{table}

\begin{table}
\begin{center}
\begin{tabular}{ccccc}
\hline\hline
\\
Process & Generator & $\rm 0.5\sqrt {\hat s}$ & $\rm \sqrt {\hat s}$ & $\rm 2.0\sqrt {\hat s}$ \\
\\
\hline
\\
$\rm \ttbar\ttbar$ & \tt MadGraph & $\rm 4.38 $ & $\rm 2.82 $ & $\rm 1.93 $ \\
                   & \tt ALPGEN   & $\rm 4.42 $ & $\rm 2.90 $ & $\rm 1.96 $ \\
                   & \tt CalcHEP  &             -             & $\rm 2.89 $ &  - \\
\\
$\rm \ttbar\bb$    & \tt MadGraph & $\rm  2.62 \times 10^{3}$  & $\rm 1.76 \times 10^{3}$ & $\rm 1.14 \times 10^{3}$  \\
                   & \tt ALPGEN   & $\rm  2.86 \times 10^{3}$ & $\rm 1.89 \times 10^{3}$ & $\rm 1.29 \times 10^{3}$ \\
                   & \tt CalcHEP  &   -   & $\rm 2.27 \times 10^{3} $ & -    \\
\\
$\rm \bb\bb$       & \tt MadGraph & $\rm  1.20 \times 10^{7}$ & $\rm 7.61 \times 10^{6}$ & $\rm 5.62 \times 10^{6}$  \\
                   & \tt ALPGEN   & $\rm  1.22 \times 10^{7}$ & $\rm 8.04 \times 10^{6}$ & $\rm 5.52 \times 10^{6} $  \\
                   & \tt CalcHEP  &   -                       & $\rm 1.15 \times 10^{7}$ &   -                        \\
\\
\hline\hline
\end{tabular}
\end{center}
\caption{ Production cross-section (fb) for different final states 
         involving four third generation quarks obtained using different
	 generators. Variation of the cross sections with QCD scale is 
         shown in the last three columns.  
 }
\label{tab:allatLHC}
\end{table}

In Table \ref{tab:allatLHC} we demonstrate the variation of the cross 
sections with the QCD scale. There is reasonable agreement between 
different generators at the same scale. As expected there is significant 
scale dependence 
in these leading order cross sections. However, in spite of this 
uncertainty  some of the background processes have cross 
sections several orders of magnitude larger than that of the signal, 
which is typically a few fb for $\rm \mglu$ near the LHC search limit
(see Table 1). Thus it is worthwhile to see the response of these 
backgrounds to our choice  of the kinematical cuts listed below as well
as to the cuts used by similar analyses \cite{lari,baer3}. We shall
subsequently show that our cuts also lead to a healthy signal size and 
control other conventional backgrounds quite effectively.

We next simulate the backgrounds listed in Table \ref{tab:allatLHC}. 
The parton level unweighted events from {\tt ALPGEN} has been interfaced 
with {\tt Pythia}. As noted earlier our final conclusion regarding 
the importance of these  backgrounds remain unchanged inspite of the 
sizable uncertainty due to the choice of scale. In order to avoid 
generation of events unlikely to pass the cuts we have imposed the following 
conditions on the parton level sub-processes. It should be noted that
these conditions are somewhat stronger than those used in Table 
\ref{tab:allatLHC}. 

\begin{itemize}
  \item $\rm\ttbar\ttbar$ events have been generated with no cut on $\rm t$.
  \item $\rm\ttbar\bb$ events have been generated with no cut on $\rm t$,
        $\rm\hat{P_{T}^{b}} > 20$ GeV and $\rm|\eta_{b}| \le 4.5$.
  \item $\rm\bb\bb$ events have been generated with
        $\rm\hat{P_{T}^{b}} > 50$ GeV and $\rm|\eta_{b}| \le 4.5$.
\end{itemize}
 
We have used the toy calorimeter simulation ({\tt PYCELL}) provided in {\tt Pythia} with the
following criteria: 
\begin{itemize}
  \item The calorimeter coverage is $\rm |\eta| < 4.5$. The segmentation is 
given
        by $\rm\Delta\eta\times\Delta\phi= 0.09 \times 0.09 $ which 
resembles
        a generic LHC detector.
  \item A cone algorithm with
        $\rm\Delta R = \sqrt{\Delta\eta^{2}+\Delta\phi^{2}} = 0.5$
        has been used for jet finding.
  \item $\rm E_{T,min}^{jet} = 30$ GeV and jets are ordered in $E_{T}$.
  \item Leptons ( $\rm \ell = e, ~\mu$ ) are selected with 
	$\rm E_T \ge 20$ GeV and $\rm |\eta| \le 2.5$.
  \item No jet should match with a hard lepton in the event.
  \item A jet with $\rm |\eta| \le 2.5$ matching with a B-hadron of 
	significant decay length has been marked {\em tagged}.
\end{itemize}

We have implemented  jet and lepton( $\rm \ell = e$ or $\rm \mu$) 
isolation using the following criteria: if there is a jet within 
$\rm \Delta R = 0.5$ and $\rm  E_{T}^{jet}/E_{T}^{\ell} \le 1.2 $, the jet 
is removed from the list of jets, else the lepton is removed from the list 
of leptons. Single b-jet tagging efficiency in $\rm t\bar{t}$ has been 
tuned to $\rm \epsilon_{b}\simeq 0.5$.

\begin{figure}[hbp]
\begin{center}
\includegraphics[width=0.95\textwidth]{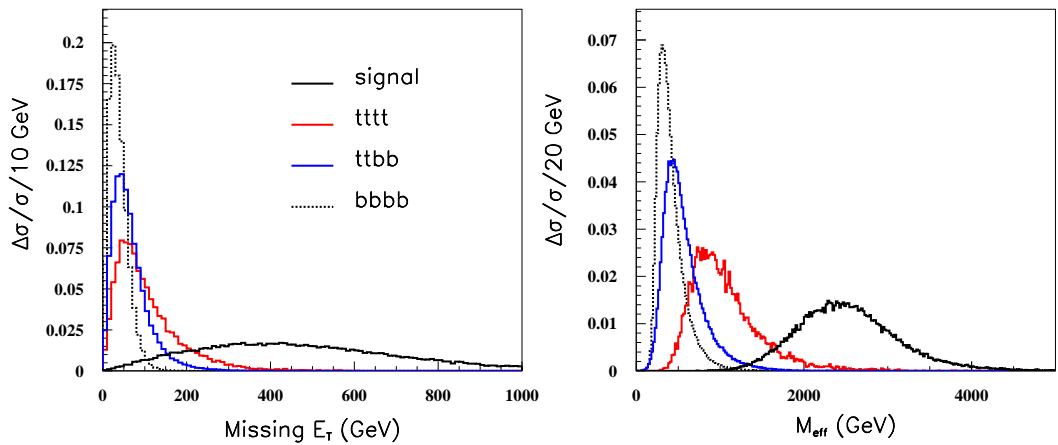}
\end{center}
\caption{ Distributions for missing transverse energy ($\rm\met$)
         (left) and effective mass ($\rm M_{eff}$) (right) for four heavy
         flavor final states and signal with $\rm\mglu = 1751.3 ~GeV$
         ( FP1, see text): in both the figures the distributions are 
         indicative of shapes only and normalized to unit cross section.}
\label{fig:4qplot}
\end{figure}
\begin{table}
\begin{center}
\begin{tabular}{lp{19mm}p{19mm}p{19mm}p{19mm}p{19mm}p{19mm}}
\hline\hline
Process & $\rm N_{jet}\ge 6$ & \rm $N_{obj}\ge 9$ & $\rm\nbtag\ge 2$ & 
$\rm P_{T}^{b1}\ge 300$ GeV & $\rm\met \ge 300$ GeV & $\rm M_{eff}\ge 2000$  GeV\\
\hline
\\
$\rm\ttbar\ttbar$ & 0.9304 & 0.5457 & 0.5946 & 0.1014 & \bf 0.0316  & \bf 0.0346 \\
$\rm\ttbar\bb$    & 0.5792 & 0.0831 & 0.4509 & 0.0265 & 0.0033    & \bf 0.0018 \\ 
$\rm\bb\bb$       & 0.2186 & 0.0064 & 0.5854 & 0.0120 & \bf 0.00004 & 0.0002 \\
\\
\hline\hline
\end{tabular}
\end{center}
\caption{Selection efficiencies for different final states involving four 
	 third generation quarks due to individual cuts have been shown. 
	 The individual cut which rejects most events for each background 
	 is shown in bold.}
\label{tab:foureff}
\end{table}
\begin{table}[hbp]
\begin{center}
\begin{tabular}{lcc|ccc}
\hline\hline
Process    & $\rm \sigma_{gen}$ & Events    & This work       & ATLAS   & 
Baer \etal\\
           &  (fb)              & generated & $\rm\residue$ & $\rm\residue$ & $\rm\residue$ \\
                     &      &                     &           &         &          \\
\hline
                     &      &                     &           &         &          \\
$\rm \ttbar\ttbar$   &$\rm  5.3$  & $\rm 1.25\times 10^{4}$ & $\rm 0.012$ & $\rm 0.093$   & $\rm  0.004$  \\
$\rm \ttbar\bb$      &$\rm 831$  & $\rm 1.40\times 10^{5}$ & $\rm 0.042$ & $\rm 0.665 $  &  $\rm  0.059$  \\
$\rm \bb\bb$         &$\rm 7260 $& $\rm 3.02\times 10^{5}$ & $\rm 0.000$ & $\rm 0.240 $  &  $\rm  0.000$   \\
                     &      &                     &           &         &          \\
\hline\hline
\end{tabular}
\end{center}
\caption{ The cross section at the generation level ($\rm \sigma_{gen}$), number of events  
          generated, and the residual cross-section ($\rm\residue$)
          after all cuts. The cross sections were computed with
          cuts on final states partons different from the ones used in Table 2
          and the default QCD scale ( $\rm Q^2= \Sigma_f(m_{f}^2 + P_{T_f}^2)$, where $\rm 'f'$      
stands for the final states partons) 
 of {\tt ALPGEN}.} 
\label{tab:fourbkgd}
\end{table}

The distributions of $\rm\met$ and $\rm M_{eff}$ for different 
final states involving four third generation quarks are shown in 
Fig. \ref{fig:4qplot}. Selection efficiencies for these background 
processes due to individual cuts are shown in Table  
\ref{tab:foureff}; the individual cut which rejects most events for 
each final state is shown in bold. The residual cross-sections for 
these backgrounds after all our selection cuts are 
shown in Table \ref{tab:fourbkgd}. As is expected from Fig. 
\ref{fig:4qplot} very little background survives.

For comparison we have computed the backgrounds corresponding to
the cuts implemented by the ATLAS collaboration \cite{lari}  
(column 5 of Table 4) and Baer \etal \cite{baer3} (column 6 of Table 4) . 
It may be  noted  that in all cases   
these  backgrounds are rather  small compared to other
backgrounds from $t \bar{t}$ production or QCD processes (see below).
This conclusion holds even if a factor of two due to 
the QCD scale uncertainty is taken into account. It may be recalled that 
the cuts in \cite{lari} are designed for relatively light gluinos in the 
focus point region. Consequently they are less severe than 
the  ones introduced  in this paper or in \cite{baer3}. As a result
some of the residual backgrounds of this type are numerically significant for  
integrated luminosities $\rm\geq  100 ~fb^{-1}$ and require some attention.

Next we turn our attention to the conventional backgrounds. First we 
shall summarize the earlier works which employed b-tagging. Mercadante 
\etal \cite{tata}  found that the dominant background comes from QCD processes 
involving light flavors and gluons only. The analysis of reference 
\cite{lari} ignored  this background arguing qualitatively that the 
requirement of 2-tagged b-jets should be adequate to reduce it to a 
negligible level. However, they studied the $\rm\bb j$ background,
which is included in the QCD background in this work
as well as in \cite{baer3} and found it to be 41\% of the  
dominant $\rm\ttbar$ background. Very recently Baer \etal
\cite{baer3} analyzed the same background with different kinematical cuts. 
 They found that this background is rather small.

It has been identified that the $\rm \ttbar$ background is the dominant one
in both \cite{baer3} and \cite{lari} and we agree with them qualitatively.  
In the former work this background was simulated by 
{\tt ISAJET} \cite{isajet}. However, in addition to other selection criteria they required 
$\rm > 7$ jets + $\rm > 2$ tagged b-jets in the final state. Since the parton 
level process has at most six jets (including two b-jets) the size 
of this background  crucially depends on the model of  parton showering   
in the generator. Our cut on $\rm N_{obj}$ is similar in nature. We shall 
briefly comment on the reliability of analyses requiring high jet multiplicity  
in the following.

We now present our estimates of these  backgrounds.
The cross-sections for QCD processes (which includes $\rm \bb$  and 
$c\bar{c}$ production) and $\rm \ttbar$ production
are quite large (see  Table 6). 
Since the selection demands a large number of jets,
$\rm \met$ and $\rm M_{eff}$ it is more
likely that events with higher $\surd{\hat{s}}$ will survive. Even then
both the above backgrounds  survive the cuts 
(mainly the one on the multiplicity of objects) because of  
initial state and final state radiation 
(ISR and FSR) and multiple interaction. On the other hand events with
lower $\surd{\hat{s}}$ have a very large share of the cross-section
and ISR and FSR help them to pass the selection. To estimate the contributions 
from the above two regions with sufficient statistics  and to avoid generating 
large number of events which will not pass the selection, we have generated 
events with certain cuts at the parton level:
\begin{itemize}
  \item $\rm \ttbar$ events have been generated in two bins,
        $\rm \hat{P_{T}^{t}} \le 600$ GeV ($\rm \ttbar$-low) and
        $\rm \hat{P_{T}^{t}} > 600$ GeV ($\rm \ttbar$-high).
  \item QCD events have also been generated in two bins,
        100 GeV $\rm \le \hat{P_{T}} \le 600$ GeV (QCD-low) and
        $\rm \hat{P_{T}} > 600$ GeV ( QCD-high) for the final state partons.
\end{itemize}

The efficiency of the individual cuts and the size of the
different backgrounds  are listed separately in Tables 5 and 6.  
The cross sections are calculated with the 
conservative choice of the QCD scale.  
It is to be noted that while the cut $\rm N_{obj} \ge 9$ is most effective 
for rejecting $\rm\ttbar$-high events with high $\rm\surd{\hat{s}}$, it is 
either the cut on $\rm\met$ or $\rm M_{eff}$ for events with low 
$\rm\surd{\hat{s}}$. The $\rm\met$ distributions before and after applying all 
other kinematical cuts are shown in Fig. 2.  
The distributions   
of $\rm N_{jet}$ and $\rm N_{obj}$ are shown in Fig. 3. It is obvious that even  
after the cut  $\rm N_{jet} >$ 7 quite a bit of background survives where as  
the cut $\rm N_{obj} >$ 9 suppresses the background more effectively. 

\begin{figure}[htbp]
\begin{center}
\includegraphics[width=0.95\textwidth]{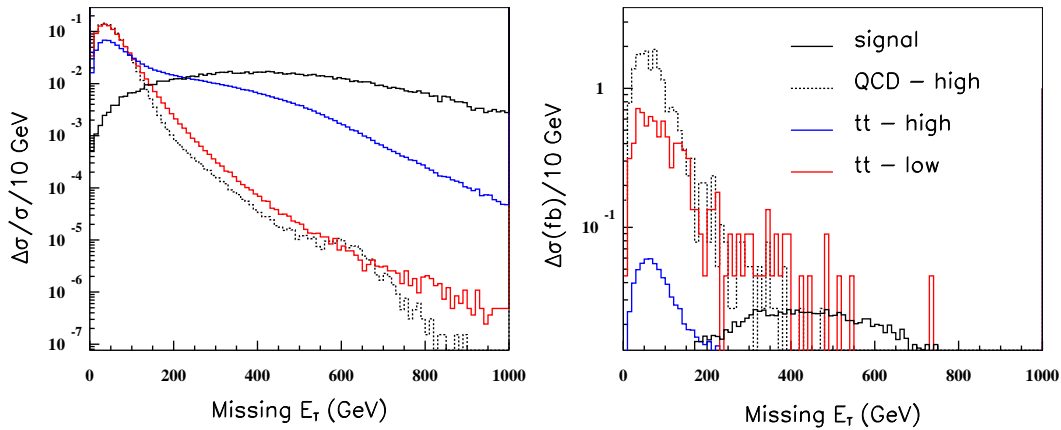}
\end{center}
\caption{ Missing transverse energy ($\rm \met$) distributions for the
         major backgrounds and the signal ($\rm\mglu = 1751.3 ~GeV$).
         In the left plot the distributions are indicative of
         the shapes only. In the right plot they have been normalized
         to the residual cross-sections after all cuts except the one
         on $\rm \met$.} 
  \label{fig:etmiss}
\end{figure}

\begin{figure}[htbp]
\begin{center}
\includegraphics[width=0.95\textwidth]{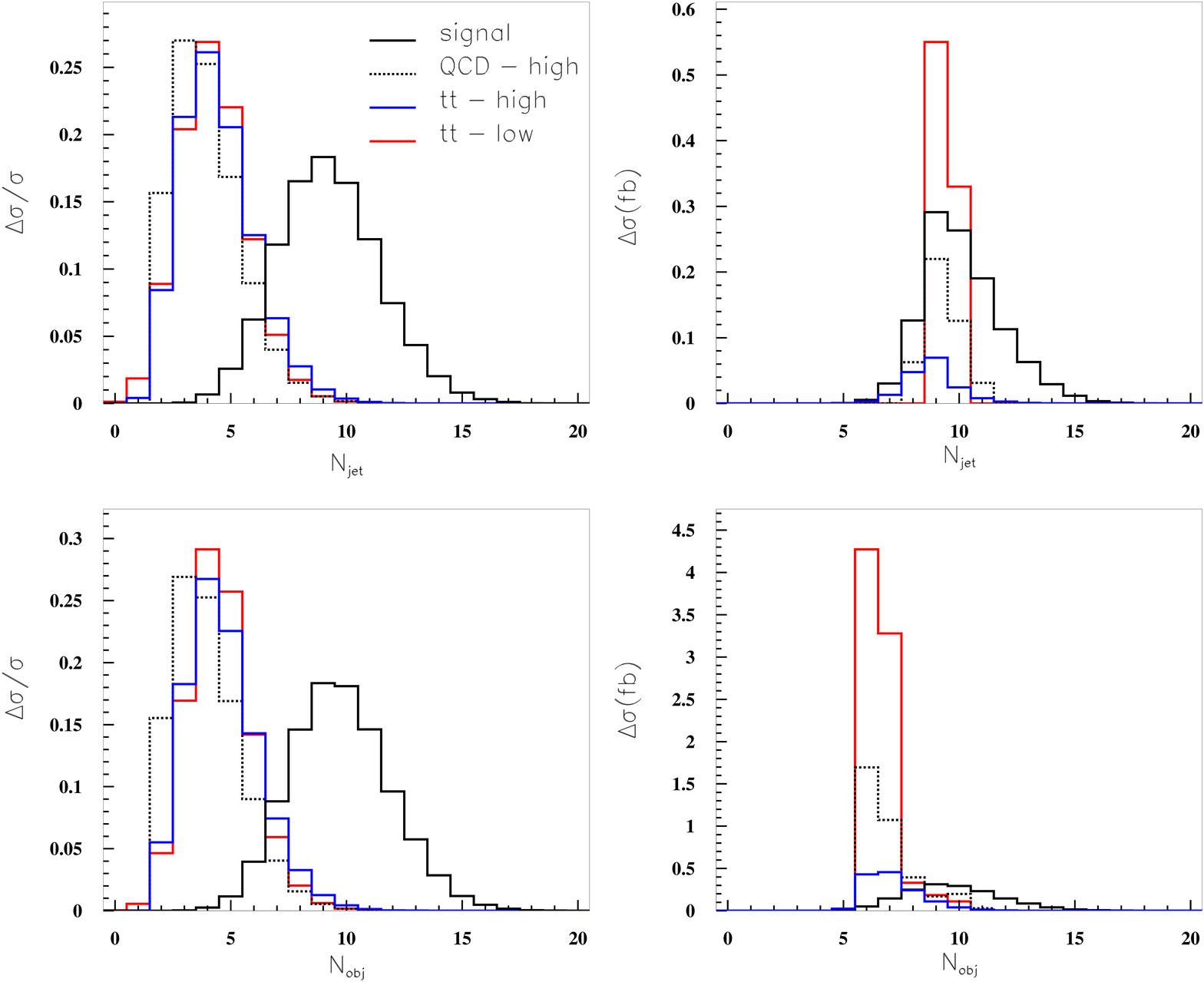}
\end{center}
\caption{The distributions of the number of jets ($\rm N_{jet}$) (top) and  
	 number of objects ($\rm N_{obj}$) (bottom) for the major 
	 backgrounds and the signal ($\rm\mglu = 1751.3 ~GeV$). 
	 The figures on the left show the distributions before any cut  
	 (normalized to unit cross section); the figures on the right 
	 show distributions normalized to residual cross-sections after  
	 all cuts except those on $\rm N_{jet}$ (top) and $\rm N_{obj}$ (bottom).
	 }
\label{fig:objects}
\end{figure}

\begin{table}
\begin{center}
\begin{tabular}{lp{19mm}p{19mm}p{19mm}p{19mm}p{19mm}p{19mm}}
\hline\hline
Process & $\rm N_{jet}\ge 6$ & $\rm N_{obj}\ge 9$ & $\rm N_{btag}\ge 2$ & $\rm P_{T}^{b1}\ge 300$ GeV & $\rm \met \ge 300$ GeV & $\rm M_{eff}\ge 2000$  GeV\\
\hline
\\
Signal           & 0.9674 & 0.7116 & 0.7899 & 0.6275 & 0.7416   & 0.8173 \\
QCD - high       & 0.1490 & 0.0074 & 0.0177 & 0.0599 & \bf 0.0018 & 0.1033 \\ 
QCD - low        & 0.0151 & 0.0001 & 0.0052 & 0.0003 & \bf 0.0000   & $\rm 1.9\times 10^{-6}$ \\
$\rm \ttbar$ - high  & 0.2318 & \bf 0.0188 & 0.4567 & 0.5853 & 0.1799 & 0.0963 \\
$\rm \ttbar$ - low   & 0.1984 & 0.0083 & 0.1490 & 0.0074 & 0.0023 & \bf 0.0014 \\
\\
\hline\hline
\end{tabular}
\end{center}
\caption{Selection efficiencies for major backgrounds due to individual cuts. 
	 The individual cut which rejects most events for 
	 each case is shown in bold.}
\label{tab:bkgd-eff}
\end{table}

\begin{table}
\begin{center}
\begin{tabular}{llccc}
\hline\hline
                     &        &                      &        &         \\
Process & $\rm \sigma_{gen}$  & Events               & 
\multicolumn{2}{c}{Residual cross-section ($\rm\residue$)} \\ \cline{4-5}
        & (fb)                & generated            & 
$\rm\nbtag \ge 2$ & $\rm\nbtag \ge 3$ \\
                     &        &                      &        &         \\
\hline
                     &        &                      &        &        \\
Signal               &        &                      &        &        \\
$\rm \mglu = 1751.3$ &  $\rm 1.20 $ & $\rm  10^{4}$ & $\rm 0.3598$ & $\rm 0.2445$  \\
$\rm \mglu = 1824.7$ &  $\rm 0.81 $ & $\rm  10^{4}$ & $\rm 0.2267$ & $\rm 0.1527$  \\
$\rm \mglu = 1950.0$ &  $\rm 0.40 $ & $\rm  10^{4}$ & $\rm 0.1445$ & $\rm 0.0987$  \\
$\rm \mglu = 2047.4$ &  $\rm 0.24 $ & $\rm  10^{4}$ & $\rm 0.0921$ & $\rm 0.0634$  \\
$\rm \mglu = 2119.8$ &  $\rm 0.16 $ & $\rm  10^{4}$ & $\rm 0.0652$ & $\rm 0.0431$  \\
\hline
                 &          &                                 &        &        \\
Background       &          &                                 &        &        \\
QCD - high       & $\rm 1.81\times 10^{5}$ & $\rm 1.08\times 10^{7}$         & $\rm 0.4362$ & $\rm 0.0504$ \\
$\rm \ttbar$ - high  & $\rm 4.14\times 10^{2}$ & $\rm 2.325 \times 10^{6}$   & $\rm 0.1675$ & $\rm 0.0205$ \\
$\rm \ttbar$ - low   & $\rm 3.70\times 10^{5}$ & $\rm 1.14\times 10^{7}$     & $\rm 0.8806$ & $\rm 0.1793$ \\
                 &          &                                 &        &        \\
\hline\hline
\end{tabular}
\end{center}
\caption{ The cross-section at the generation level ($\rm \sigma_{gen}$) , number of events
          generated, and residual cross-section
          ($\rm \residue$)
          after all cuts are shown. $\rm \sigma_{gen}$ is calculated for $\rm Q^2 = \hat{s}$}.  
\label{tab:bkgd-xsec}
\end{table}

\begin{figure}
\begin{center}
\includegraphics[width=0.95\textwidth]{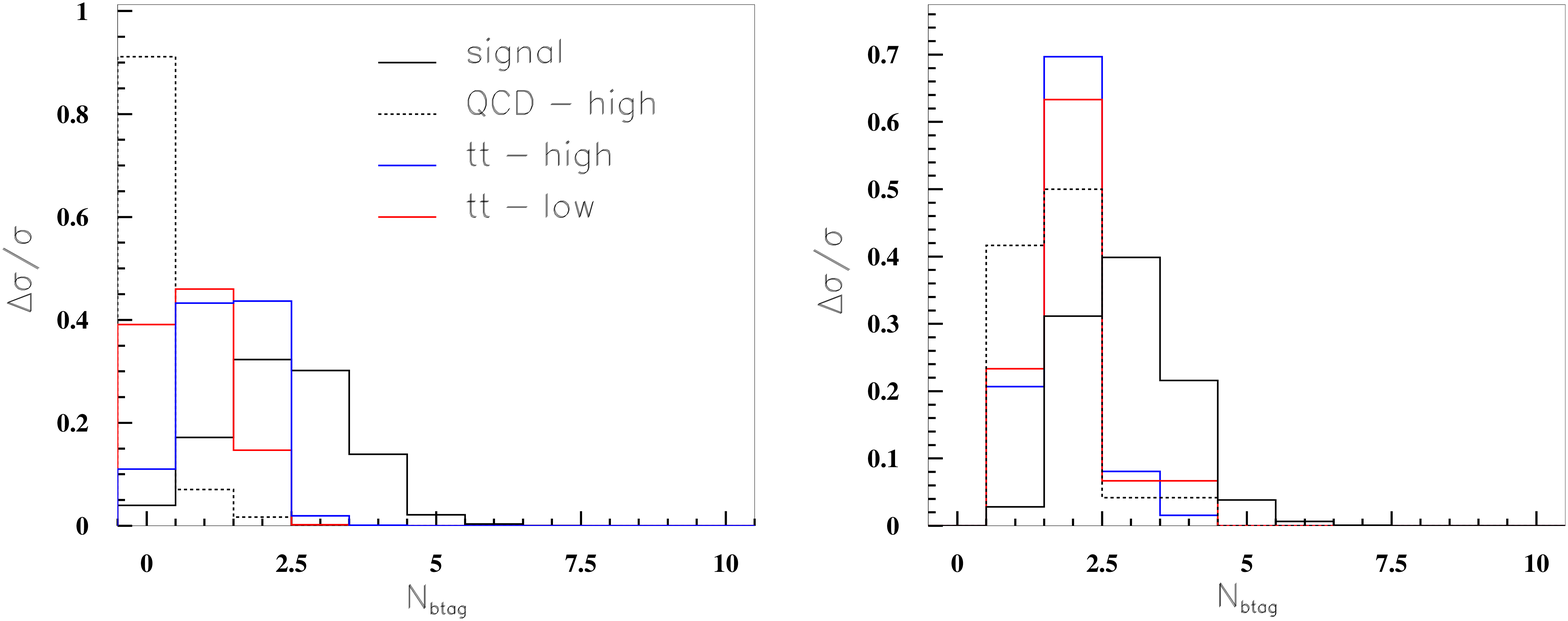}
\end{center}
\caption{Distributions for number of b-tagged jets ( $\rm\nbtag$ ) are 
shown:
         (left) before any selection criteria has been applied and (right)
         after other selection criteria have been applied. In both the
         distributions are indicative of shapes only.}
\label{fig:btag}
\end{figure}

We do not impose any extra strong hardness cut on a jet which is counted 
as  an acceptable object. It is quite plausible that in a real life 
situation such jets indeed  arise from showering. However as noted before 
the number of such jets in an event crucially depends on the parton 
showering model in the generator. Therefore the reliability of the 
background estimate both in this work as well as in \cite{baer3}  can  
not be guaranteed  until the showering model is validated using LHC data. 
In the next paragraph we briefly summarize a study \cite{mangano} which 
indicates that the $\rm \ttbar$ background estimated as above may not 
be wide off the mark. 

It is certainly desirable to supplement the lowest order estimates of
the background by higher order matrix element calculations.
However, care must be taken to avoid double counting of jets which may 
arise in a lower order calculation due to hard emissions during shower 
evolution as well as from a higher order matrix element. A proper 
matching of matrix elements and shower evolution is, therefore, called 
for. In reference \cite{mangano} the lowest order cross section of 
$\rm \ttbar$ production was computed. Next the combined cross section 
of the processes $\rm \ttbar$(exclusive), $t\bar{t}j$(exclusive), 
$t\bar{t}jj$(exclusive)and  $t\bar{t}jjj$(inclusive), where 
j stands for a light quark or gluon, were obtained after matrix 
element matching( see Table 1 and Table 2 of \cite{mangano}).
The two results agree nicely both for Tevatron and LHC energies.
A similar agreement is found  for various distributions of the 
produced $t\bar{t}$ system computed by the two different  ways. 
It is these  distributions which after all determine the efficiencies 
of various cuts. Thus one may conclude that at least the lowest order 
$\rm \ttbar$ background including parton showering which is the dominant
background for the signal under study is fairly reliable.  
   
The  case of the pure light flavor QCD background is rather  similar. 
The required final state with nine objects from this 2 $\ra$ 2 process can 
arise solely due to parton showering and, therefore, model dependent 
to some extent. Simulating the lowest order QCD processes 
in {\tt Pythia} we find that the size of this background 
is smaller than the $\rm \ttbar$ background but is non-negligible.
However the entire contribution comes from QCD-high events
$\rm \hat{P_{T}} > 600$ GeV. After generating
15 million QCD-low (100 GeV $\rm \le \hat{P_{T}} \le 600$ GeV)
events we found that no event survive the cuts.

We now compare our results with other calculations which required
two or more tagged b-jets in the final state. A 
comparison with
\cite{lari} may not be meaningful as their cuts are designed for
the search of a relatively light gluino ($\rm \mglu$ = 857.0).
In fact with their cuts we find that the survival probability of
the $\rm \ttbar$ background is twice as large as our estimate and will 
overwhelm the gluino signal for $\rm \mglu\approx$ 2 TeV, which is 
roughly the LHC reach in the FP region, having  a cross section not 
exceeding a few fb's.
 
The $\rm\ttbar$ background after all cuts estimated by Baer 
\etal \cite{baer3} is 1.5 fb with 'a tiny' contribution from  QCD. On the other 
hand we get a smaller $\rm \ttbar$ background while the QCD background 
is non-negligible. Thus one may conclude that our cuts are more efficient 
in reducing the former background while the cuts of \cite{lari} 
may be better for eliminating the QCD backgrounds. A more detailed 
comparison is not possible since we donot know the QCD scales for  
different cross sections in \cite{baer3}. 

It is probably a numerical conspiracy that the total background
($\rm \ttbar$ + QCD) computed by the two groups agree rather nicely.
However, the fact that two groups using different event generators
and different sets of kinematical cuts get comparable results
at least indicates that the major backgrounds obscuring the
search for focus point SUSY at LHC are fairly well understood.

We next present the signal size for different gluino masses. The 
efficiencies of the  cuts used in this paper for the scenario FP1 are
in Table 5 while the signal size for the conservative scale choice 
in different scenarios after 
all cuts are in Table \ref{tab:bkgd-xsec}. The signal cross sections 
should be multiplied by a factor of 2 if the scale $\rm \mglu/$2 is considered 
(see Table 1). The $\rm \met$, $\rm N_{jet}$ and $\rm N_{obj}$
distributions are in Figs. 2 and 3. Comparing these with the background
cross sections and taking the QCD scale uncertainty into account it is 
clear that the gluino mass reach at the LHC  cannot be pin pointed as yet. 
Nevertheless  the following comments are in order.

Using the optimistic scale choice (= $\rm \mglu/$2) one finds that for FP1(FP3)
(corresponding to $\rm \mglu$ = 1751.3 (1950)) 
S/$\sqrt{B}$=10.0 (4.0)  with an integrated luminosity 300 fb $^{-1}$, 
where S and B are the number of signal and background events respectively. Of 
course the S/$\sqrt{B}$ ratio may be modestly increased by optimizing  the
cuts \cite{tata}. 
On the other hand  the LO signal cross section at the scale $\rm \mglu/$2 
mimics the NLO cross section \cite{zerwas}. In view of this the  next to 
leading order
$t\bar{t}$ cross section (about a factor of 2 larger than the cross 
section in Table \ref{tab:bkgd-xsec} ) may  give a more reliable background 
estimate. This
will further suppress the  S/$\sqrt{B}$  ratio by a factor of $\sqrt{2}$.  
 For the conservative scale choice on the
other hand the above ratios  will be suppressed by a factor of two.
On the whole it seems unlikely that
using the selection criteria based on 2 or more tagged b-jets as is
usually required, the gluino mass reach at LHC
can be pushed far beyond 1.8 TeV with certainty.  

We now impose a more stringent b-identification criteria as suggested 
in \cite{utpal} and  require $\geq$ 3 b-jet tags keeping all other 
selection criteria listed above  unchanged.
The residual signal and the backgrounds are shown in the last column of 
Table 6. Even the conservative scale choice now yields S/$\sqrt{B}$
ratio 8.5 (3.4) for $\rm \mglu$ = 1751.3 (1950). For the optimistic scale 
choice the above ratios are enhanced approximately by a factor of $\sqrt{2.0}$
after doubling the background which should reflect the theoretical 
uncertainties due to possible NLO corrections etc.
In view of the possibility of further improving this ratio by 
optimizing the selection criteria,
one can hope that irrespective of the uncertainty due to the choice of 
scale the reach in $\rm \mglu$ at LHC is close to 2.0 TeV.

\section{Conclusion}

If supersymmetry in the focus point region is realized in nature it may 
lead to spectacular signals at the LHC via gluino pair production \cite
{utpal,baer1,tata,cms,belyaev,lari,baer3}. We propose a new set of criteria
for selecting(rejecting) the signal( background) (see section 2).

We have extended the estimation of the backgrounds which 
may hinder the search for supersymmetry in the focus point region.
In particular we have simulated and analyzed the potentially dangerous
backgrounds from different processes involving four heavy flavor quark
production which have so far been neglected. Some of these processes have 
cross sections several orders of magnitude larger than that of the signal. 
Moreover these backgrounds are rich in b-jets and cannot be reduced 
by b-tagging which is recommended as one of the most effective 
tools for removing other SM backgrounds \cite{utpal,tata,baer3}. 

Using rather  stringent cuts (see section 2) 
appropriate for detecting relatively heavy gluinos with masses near the 
kinematic reach of the LHC, we have shown that these background are highly 
suppressed(see Table 4). On the other hand for less severe cuts often 
employed  for relatively light gluinos in the focus point region (see  
for example \cite{lari}), these backgrounds may be  small but numerically 
significant and call for some attention.  

We have also simulated the conventional SM backgrounds estimated by other 
groups. We qualitatively agree with the observation \cite{lari,baer3}
that $\rm \ttbar$ events are  the most severe background to the gluino signal.
In view of various theoretical uncertainties in the production cross 
sections
the LHC reach in $\rm \mglu$ cannot
be pinpointed as yet. This is especially so if the selection criteria
is restricted to  
$\geq$ 2 tagged b-jets which is the usual strategy. Requiring 3 or more 
tagged b-jets in the signal the gluino mass reach at LHC
is close to 2.0 TeV even with a modest production cross section corresponding 
to a conservative choice of the QCD scale.   

We look forward to improved b-tagging efficiencies along 
with improvement in the theoretical understanding (e.g., the inclusion of 
matrix element matching) of the light flavor QCD  and other backgrounds  
for a better handle on gluino searches in the FP scenario.  

{\bf Acknowledgement}: SPD acknowledges postdoctoral fellowships from 
Bundesministerium f\"ur Bildung und Forschung (BMBF) Projekt under Contract No. 05HT6PDA 
and Harish Chandra Research Institute where part of this work  
was done. AD and MM acknowledge support by the DST, India
(project SR/S2/HEP-18/2003). MM and SM acknowledge support by the DST,
India (project SP/S2/K-25/96-VI). The authors thank M. Drees, H. Driener,  
M.L. Mangano and S. Mrenna for many useful suggestions and discussions.


\begin{thebibliography}{99}
\bibitem{mtop}The recent world average value for the top quark 
mass is $\mtop$ =170.9 GeV \cite{mtop1}. 
We have, however, chosen the reference value $\mtop$ =175 GeV . 
This facilitates the  comparison 
with other published studies.
\end{thebibliography}
\end{document}